\documentclass{article}
 
 \usepackage{mathptmx}
\usepackage{graphicx} % more modern
\usepackage[nooneline,tight]{subfigure}%\usepackage{subcaption}
\usepackage{amsfonts}
\usepackage{amsmath}
\usepackage{commands}
\usepackage{url}
\usepackage{color}

\usepackage{natbib} 

\usepackage{algorithm}
\usepackage{algorithmic}

\usepackage{hyperref}

\usepackage[accepted]{icml2014}

\newcommand{\ngram}{$n$-gram}

\newcommand{\dims}{D}
\newcommand{\certain}{deterministic}
\newcommand{\Certain}{Deterministic}
\newcommand{\token}{\alpha}
\newcommand{\tokenIdx}{t} 
\newcommand{\tokenLen}{T} 
\newcommand{\tokens}{\boldsymbol{\token}}

\newcommand{\node}{n}
\newcommand{\nodeIdx}{i} 
\newcommand{\nodeLen}{N} 

\newcommand{\oneLatent}{h}
\newcommand{\latent}{\boldsymbol{h}}
\newcommand{\latentIdx}{j} 
\newcommand{\latentLen}{H} 
\newcommand{\latents}{\latent_{0:\nodeLen}}

\newcommand{\varfeat}{\boldsymbol{v}}
\newcommand{\varfeatIdx}{u}
\newcommand{\varfeatLen}{V}

\newcommand{\stack}{S}

\newcommand{\children}{\boldsymbol{C}}
\newcommand{\representation}{R}

\newcommand{\bias}{b}
\newcommand{\contextmatrixcon}{W^{con}}
\newcommand{\contextmatrixch}{W^{ch}}

\newcommand{\default}{def}

\newcommand{\hmmStateSize}{K}

\newcommand{\OurModelLong}{Log-bilinear Tree-Traversal}
\newcommand{\OurModelShort}{LTT}

\DeclareMathOperator{\prob}{p}
\DeclareMathOperator{\energy}{E}

\DeclareMathOperator{\repfn}{R}
\DeclareMathOperator{\biasfn}{b}
\DeclareMathOperator{\reversed}{\textsc{Reversed}}

\newcommand\blfootnote[1]{%
  \begingroup
  \renewcommand\thefootnote{}\footnote{#1}%
  \addtocounter{footnote}{-1}%
  \endgroup
}
\icmltitlerunning{Structured Generative Models of Natural Source Code}

\begin{document} 

\twocolumn[
\icmltitle{Structured Generative Models of Natural Source Code}

% It is OKAY to include author information, even for blind
% submissions: the style file will automatically remove it for you
% unless you've provided the [accepted] option to the icml2013
% package.
\icmlauthor{Chris J. Maddison\textsuperscript{\dag}}{cmaddis@cs.toronto.edu}
\icmladdress{University of Toronto}
\icmlauthor{Daniel Tarlow }{dtarlow@microsoft.com}
\icmladdress{Microsoft Research}

% You may provide any keywords that you 
% find helpful for describing your paper; these are used to populate 
% the "keywords" metadata in the PDF but will not be shown in the document
\icmlkeywords{}

\vskip 0.3in
]

\begin{abstract} 
We study the problem of building generative models of
\emph{natural} source code (NSC); that is, source code written 
by humans and meant to be understood by humans.
%---it is important, amenable to machine learning methods, and
%the unique nature of natural source code presents new challenges
%that can drive forward fundamental machine learning research.
%
%Our primary contribution is to describe a family of generative
%models for NSC that have two key properties: First, they
%incorporate both sequential and hierarchical structure.
Our primary contribution is to describe new 
generative models that are tailored to NSC.
The models are based on probabilistic context free grammars (PCFGs) and
neuro-probabilistic language models \cite{MnihTeh2012}, which are extended to
incorporate additional source code-specific structure.
%This allows, for example, maintaining the set of variables
%that are currently in scope throughout the generative process.
%models for NSC that have two key properties: First, they
%Second, they are capable of integrating closely with a compiler, which allows
%leveraging compiler logic and abstractions
%when building structure into the model. 
%We also develop
%an extension that includes more complex structure,
%refining how the model generates identifier tokens based on
%what variables are
%currently in scope.  
%Our models are highly structured, but the structure is chosen so
%as to support efficient learning and evaluation of data log probabilities.
These models can be efficiently trained on a corpus of source code
and outperform a variety of less structured baselines in terms of predictive log likelihoods
on held-out data.
%
%introduce the problem from a machine learning
%perspective then develop new models that (a) simultaneously capture
%the hierarchical and sequential structure present in NSC, and (b)
%incorporate complex deterministic functions, like those that are implemented
%inside of a compiler for the purposes of type-checking and determining
%which variables are in scope.  Of particular interest is the combination
%of deterministic mappings that expand the space paired with log-bilinear
%parameterizations to mitigate data fragmentation.
%We show that these models can be learned efficiently using
%EM and noise contrastive estimation (NCE),
%and we show empirically that including appropriate structure significantly
%improves log probabilities of generating held out programs.
\vspace{-8pt}
\end{abstract}

%%%%%%%%%%%%%%%%%%%%%%%%%%%%%%%%%%%%%%%%%%%%%%%%%%%%%%%%%%%%%%%%%%%%%%%%%%%%%%%%%%%%%%%%%
%%%%%%%%%%%%%%%%%%%%%%%%%%%%%%%%%%%%%%%%%%%%%%%%%%%%%%%%%%%%%%%%%%%%%%%%%%%%%%%%%%%%%%%%%
\vspace{-5pt}
\section{Introduction}
\vspace{-5pt}

Source code is ubiquitous, and a great deal of human effort goes into
developing it.  An important goal is to develop tools that make the
development of source code easier, faster, and less error-prone, and
to develop tools that are able to better understand pre-existing
source code.  To date this problem has largely been studied outside
of machine learning. Many problems in this area
do not appear to be well-suited to current machine learning technologies.
  Yet, source code is some of
the most widely available data with many public online
repositories. Additionally, massive open online courses (MOOCs)
have begun to collect source code homework assignments from tens of
thousands of students \cite{huang2013syntactic}.  
At the same time, the software engineering
community has recently observed that it is useful to think of source
code as natural---written by humans and meant to be understood by
other humans \cite{hindle2012naturalness}. 
This \emph{natural source code} (NSC) has a great deal of statistical
regularity that is ripe for study in machine learning.

The combination of these two observations---the availability of data,
\blfootnote{\textsuperscript{\dag}Work done primarily while author was an intern at Microsoft Research.}
and the presence of amenable statistical structure---has opened up
the possibility that machine learning tools could become useful in various tasks
related to source code.  At a high level, there are several potential
areas of contributions for machine learning. First, code editing
tasks could be made easier and faster.
Current autocomplete suggestions rely primarily on heuristics
developed by an Integrated Development Environment (IDE) designer.
With machine learning methods, we might be able to offer much improved
completion suggestions by leveraging the massive amount of source code
available in public repositories.  Indeed,
\citet{hindle2012naturalness} have shown that even simple \ngram~models are
useful for improving code completion tools, and \citet{nguyen2013statistical}
have extended these ideas.  Other related applications 
include finding bugs \cite{kremenek2007factor}, mining and suggesting API usage patterns
\cite{bruch2009learning,nguyen2012graphbased, wang2013mining}, as a
basis for code complexity metrics \cite{allamanis2013mining},
and to help with enforcing coding conventions \cite{allamanis2014learning}.
Second, machine learning might open up whole new applications such as
%the possibility of new source
%code-related applications that would not otherwise be possible.  Here,
%we believe it productive to view source code as a new type of
%data for research on structured prediction.  New potential
%applications inspired by this view include include 
automatic translation between programming languages, automatic code
summarization, and learning representations of source code for
the purposes of visualization or discriminative learning.
Finally, we might hope to leverage the large amounts of
existing source code to learn improved priors over programs for use in
programming by example
\cite{halbert1984programming,gulwani2011automating} or other
program induction tasks.

One approach to developing machine learning tools for NSC is to
improve specific one-off tasks related to source code.  Indeed, much
of the work cited above follows in this direction.  An alternative,
which we pursue here, is to develop a generative model of source code
with the aim that many of the above tasks then become different forms
of query on the same learned model (e.g., code completion is
conditional sampling; bug fixing is model-based denoising;
representations may be derived from latent variables
\cite{hinton2006reducing} or from Fisher vectors
\cite{jaakkola98exploiting}).  We believe that building a
generative model focuses attention on the challenges that source code
presents. It forces us to model all aspects of the code, from the high
level structure of classes and method declarations, to constraints
imposed by the programming language, to the low level details of how
variables and methods are named.  We believe building good models of NSC 
to be a worthwhile modelling challenge for machine learning research to
embrace.

%\TODO{explain that generative modelling requires explaining everything about
%code. so we need to capture things like programming language constraints, 
%but also stylistic things like variable names and code patterns.}

%\commentout{\TODO{not sure I like this... too jargony. }
%In building a generative model of NSC, the structural constraints imposed by 
%a language specification are one of the primary challenges. Yet existing tools
%for manipulating source code do an excellent job of capturing these
%constraints; for example, Microsoft's IntelliSense
%\citet{IntelliSense} makes code completion suggestions that are aware
%of the kind of token that is expected at the current cursor position,
%and it is aware of a great deal of context, such as which variables
%are currently in scope.  So while there is potentially much to be
%gained by effectively leveraging large amounts of data, it is likely
%that significant progress will require incorporating these source
%code-specific structural constraints into our models.
%This is one of the primary challenges that we address in this work.
%}

\begin{figure}
%\footnotesize{
%\begin{verbatim}
%for (int i = 1000; for(int i = 0; for < i = i; 
%  for ] i[i] = i;
%\end{verbatim} 
%}
%\centering
%{\bf (a) Log-bilinear $n$-gram Model}
%\vspace{10pt}
\footnotesize{
\begin{verbatim}
for (m[canplace] = -0; i[0] <= temp; m++) {
   co.nokori = Dictionary;
   continue;
}
\end{verbatim} 
}
\vspace{-10pt}
\centering
{\bf (a) Log-bilinear PCFG}

\vspace{10pt}
\footnotesize{
\begin{verbatim}
for (int i = words; i < 4; ++i) {
   for (int j = 0; j < i; ++j) {
      if (words.Length % 10 == 0) { 
         Math.Max(j+j, i*2 + Math.Abs(i+j)); 
      }
   }
}
\end{verbatim} 
}
\vspace{-20pt}
\centering
{\bf (b) Our Model}

\caption{Samples of \code{for} loops generated by learned models. Our
model captures hierarchical structure
 and other patterns of variable usage
and local scoping rules. Whitespace edited to improve readability.}
\label{fig:intro_fig}
\end{figure}

In \secref{overview} we introduce notation and motivate the requirements of our models---they must capture the sequential
and hierarchical nature of NSC, naturalness, and code-specific
structural constraints.  In Sections \ref{modeldesc} and
\ref{sec:extending} we introduce \OurModelLong{} models
(\OurModelShort{s}), which combine natural language processing models
of trees with log-bilinear parameterizations, and additionally
incorporate compiler-like reasoning.  In
\secref{sec:inference_and_learning} we discuss how efficiently to learn 
these models, and in
\secref{sec:experiments} we show empirically that they far outperform
the standard NLP models that have previously been applied to source
code.  
As an introductory result, \figref{fig:intro_fig} shows samples generated 
by a Probabilistic Context Free Grammar (PCFG)-based model (\figref{fig:intro_fig} (b)) versus samples generated
by the full version of our model (\figref{fig:intro_fig} (b)).
%captures far more structure, both natural and structural, than existing
%models that have been applied to source code.
Although these models apply to any common imperative
programming language like C/C++/Java, we focus specifically on C\#.
This decision is based on (a) the fact that large quantities of data
are readily available online, and (b) the recently released Roslyn C\#
compiler \cite{ROSLYN} exposes APIs that allow easy access to a rich
set of internal compiler data structures and processing results.

\vspace{-5pt}
\section{Modelling Source Code}
\vspace{-5pt}
\label{overview}
In this section we discuss the challenges in building a generative
model of code. In the process we motivate our choice of representation and model and introduce terminology that will
be used throughout.

\begin{figure}[t]
\begin{center}
\centerline{\includegraphics[width=.8\columnwidth]{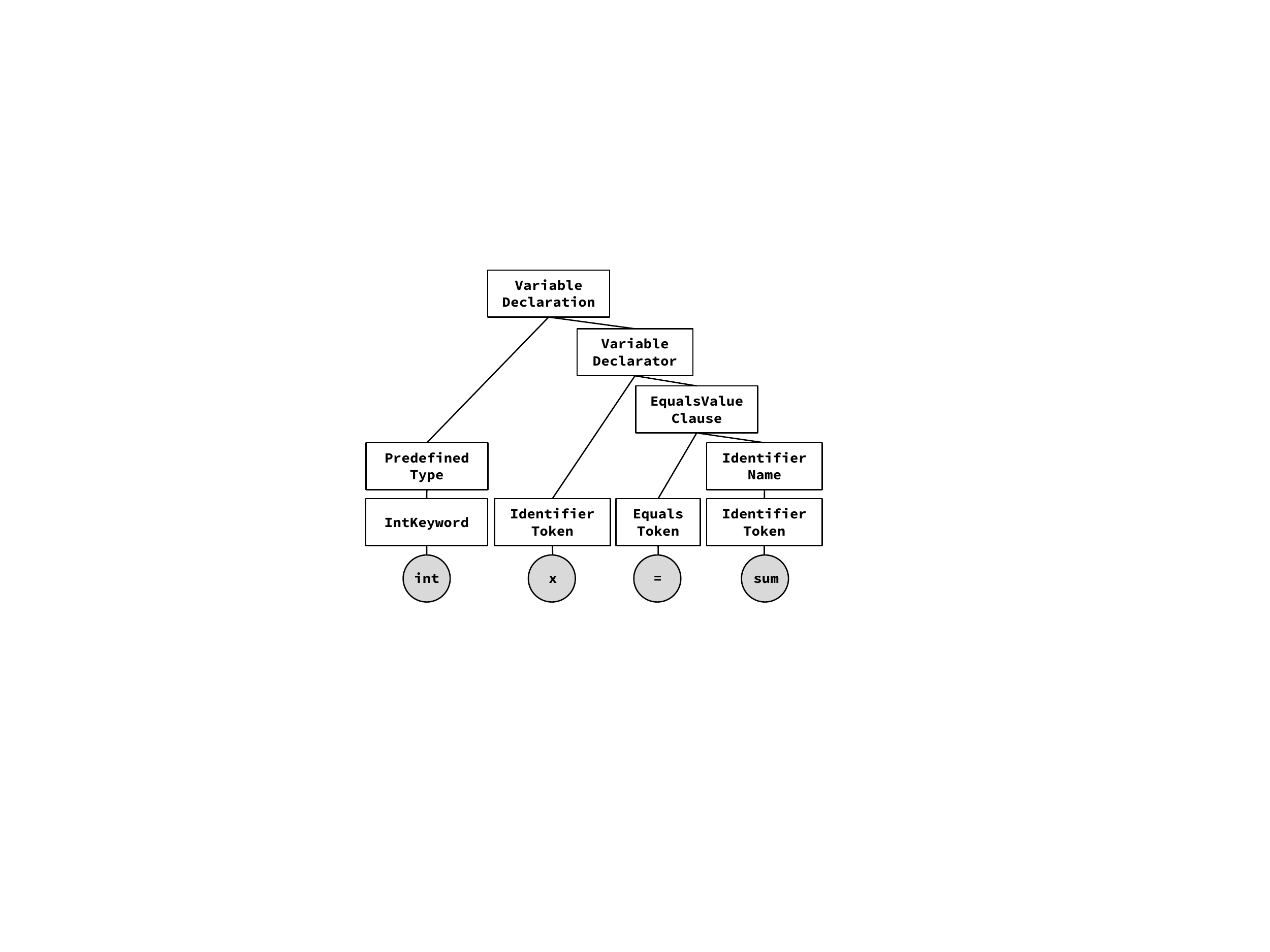}}
\caption{Example Roslyn AST. Rectangles are internal nodes and shaded circles are tokens.}
\vspace{-20pt}
\end{center}
\label{fig:ast}
\end{figure}

\textbf{Hierarchical Representation. }
The first step in compilation is to lex code into a sequence of
\emph{tokens}, $(\token_{\tokenIdx})_{\tokenIdx=1}^{\tokenLen} =
\tokens$. Tokens are strings such as ``\code{sum}'', ``.'', or
``\code{int}'' that serve as the atomic syntactic elements of a
programming language. However, representing code as a flat sequence
leads to very inefficient descriptions. For example, in a C\#
\code{for} loop, there must be a sequence containing the tokens
\code{for}, \code{(}, an initializer statement, a condition
expression, an increment statement, the token \code{)}, then a body
block.
%the body of a \code{for} loop can contain nested loops, 
A representation that is fundamentally flat cannot compactly
represent this structure, because \code{for} loops
can be nested. Instead, it is more efficient to use the
hierarchical structure that is native to the programming
language. Indeed, most source code processing is done on tree
structures that encode this structure. These trees are called abstract
syntax trees (ASTs) and are constructed either explicitly or
implicitly by compilers after lexing valid sequences of code. The leaf
nodes of the AST are the tokens produced by the lexer. The
\emph{internal nodes} $\{\node_{\nodeIdx}\}_{\nodeIdx = 1}^{\nodeLen}$
are specific to a compiler and correspond to expressions, statements
or other high level syntactic elements such as \code{Block} or
\code{ForStatement}. The \emph{children tuple} $\children_{\nodeIdx}$ of an internal node $\node_{\nodeIdx}$ is a
tuple of nodes or tokens.
An example AST is shown in
\figref{fig:ast}. Note how the \code{EqualsValueClause}
node has a subtree corresponding to the code \code{= sum}.
Because many crucial properties of the source code
can be derived from an AST, they are a primary data structure used
to reason about source code. For example, the tree structure is enough to
determine which variables are in scope at any point in the program. For this reason we choose the AST as the representation for source code and consider generative models that define distributions over ASTs.
%
%Source code begins as one long string. The first task a compiler
%performs is to lex the code into  Given $\tokens$,
%a compiler constructs an AST. The AST is a tree of \emph{nodes}
%that represents the
%syntactic structure of code and that serves as an intermediate data
%structure used extensively by semantic analysis tools.
%sequence of tokens (and previous work has done so), a flat representation is a
%difficult representation to build on. Basic constraints become
%difficult to encode;   
%
%The AST contains the hierarchical structure of the
%source code, and it also labels  each internal nodes
%according to the code construct contained in the subtree of decendants under the
%node. Leaf nodes correspond to \emph{tokens}, which are strings representing
%the atomic syntactic elements of the program.

\textbf{Modelling Context Dependence. }
A PCFG seems like a natural choice for modelling ASTs. PCFGs generate ASTs from the
root to the leaves by repeatedly sampling children tuples given
a parent node. The procedure recurses until all leaves are tokens producing nodes $\node_{\nodeIdx}$ in a depth-first traversal order and sampling children tuples independently of the rest of the tree. Unfortunately this independence assumption produces a weak model;
\figref{fig:intro_fig} (a) shows samples from such a model.
While basic constraints like matching of parentheses and braces are
satisfied, most important contextual dependencies are lost. For example, identifier names (variable and method names) are 
drawn independently of each other given the internal nodes of the AST,
leading to nonsense code constructions.

The first source of context dependence in NSC comes from the naturalness
of software. People have many stylistic habits that significantly
limit the space of programs that we might expect to see.
For example, 
when writing nested for loops, it is common to name the
outer loop variable \code{i} and the inner loop variable \code{j}.
The second source of context dependence comes from additional constraints inherent
in the semantics of code. Even if the syntax is context free, the fact that a program conforms to the grammar does
not ensure that a program compiles. For example, variables
must be declared before they are used.

%when writing nested for loops, it is common to name the
%outer loop variable \code{i} and the inner loop variable \code{j}.
%People tend not to nest blocks too deeply.

%Our approach to dealing with this sort of dependence will be to
%maintain a learned latent representation of context (a real-valued
%vector) at each node in an AST. The context vector will be a function
%of ancestry in the AST and of previous tokens, and it will modulate
%the decisions that are made during the generative procedure. Details
%of the representation are given in \secref{subsec:logbilinear_parameterization}.

%,
%but a program conforming to a context free grammar does not ensure that
%this constraint is satisfied.
%To incorporate this structure into our model, we augment the generative
%model with data structures that would be used in a compiler to check
%whether a program is valid, so that we can ensure that certain constraints
%are satisfied in the programs that the model generates. 
%In \secref{sec:extending} we describe the details, showing
%how to incorporate information about the set of variables that are
%currently in scope into the model.

Our approach to dealing with dependencies beyond what a PCFG can represent
will be to introduce
\emph{traversal variables} $\{\latent_{\nodeIdx}\}_{\nodeIdx = 0}^{\nodeLen}$ that evolve
sequentially as the nodes $\node_{\nodeIdx}$ are being generated. Traversal variables
modulate the distribution over children tuples by maintaining a representation of context that depends on the state of the AST generated so far.

%Key throughout this all is that we desire to make modelling choices
%that ensure that we can efficiently sample from the model, evaluate
%the log probability of programs under the model, and efficiently
%learn model parameters from a corpus of source code. Details of
%the efficient computations are given in \secref{sec:inference_and_learning}.

%we provide a brief review of compiler organization and concepts that will
%be useful, as well as a brief discussion of specifics related to the Roslyn compiler.s
%%%%%%%%%%%%%%%%%%%%%%%%%%%%%%%%%%%%%%%%%%%%%%%%%%%%%%%%%%%%%%%%%%%%%%%%%%%%%%%%%%%%%%

\vspace{-5pt}
\section{\OurModelLong{} Models}
\vspace{-5pt}
\label{modeldesc}
\begin{algorithm}[t]
{\footnotesize
\caption{Sampling from \OurModelShort{}s.
%, the tree can be constructed by drawing edges between parents $\node_{\nodeIdx}$ and children $\children_{\nodeIdx}$ as they are sampled.
}
\label{genproc}
\begin{algorithmic}[1]
\STATE initialize empty stack $\stack$
\STATE sample $(\node, \latent_{0}) \sim \prob(\node, \latent_{0})$
\STATE push $\node$ onto $\stack$
\STATE $(\nodeIdx, \tokenIdx) \gets (1, 1)$
\WHILE{$\stack$ is not empty}
\STATE pop the top node $\node$ from $\stack$
\IF {$\node$ is an internal node}
\STATE $\node_{\nodeIdx} \gets \node$
\STATE sample $\latent_{\nodeIdx} \sim \prob(\latent_{\nodeIdx}  \given \latent_{\nodeIdx -1})$ 
\STATE sample $\children_{\nodeIdx} \sim \prob(\children_{\nodeIdx} \given \node_{\nodeIdx}, \latent_{\nodeIdx})$
\STATE push $\node$ for $\node \in \reversed(\children_{\nodeIdx})$ onto $\stack$
\STATE $\nodeIdx \gets \nodeIdx + 1$
\ELSE
\STATE  $\token_{\tokenIdx} \gets \node$
\STATE $\tokenIdx \gets \tokenIdx + 1$
\ENDIF
\ENDWHILE
\end{algorithmic}
}
\label{alg:sample_ltt}
\end{algorithm}

\OurModelShort{s} are a family of probabilistic models that generate ASTs in a 
depth-first order (Algorithm \ref{genproc}). First, the stack is initialized and the root is pushed (lines 1-4). Elements are popped from the stack until it is empty. If an internal node $\node_{\nodeIdx}$ is popped (line 6), then it is expanded into a children tuple and its children are pushed onto the stack (lines 10-11). If a token $\token_{\tokenIdx}$ is popped, we label it and continue (line 14). This procedure has the effect of generating nodes in a depth-first order.
In addition to the tree that is generated in a recursive fashion, traversal variables $\latent_{\nodeIdx}$ are updated whenever an internal node is popped (line 9). Thus, they traverse the tree, evolving \emph{sequentially}, with each $\latent_{\nodeIdx}$ corresponding to some partial tree of the final AST. This sequential view will allow us to exploit
context, such as variable scoping, at intermediate
stages of the process (see \secref{sec:extending}).

Algorithm \ref{genproc} produces a
sequence of internal nodes $(\node_{\nodeIdx})_{\nodeIdx =
1}^{\nodeLen}$, traversal variables $(\latent_{\nodeIdx})_{\nodeIdx =
0}^{\nodeLen}$, and the desired tokens $\{\token_{\tokenIdx}\}_{\tokenIdx = 1}^{\tokenLen}$.
It is defined by three
distributions: (a) the prior over the root node and traversal
variables, $\prob(\node, \latent)$; (b) the distribution over
children tuples conditioned on the parent node and $\latent$, denoted
 $\prob(\children \given \node,\latent)$; and
(c) the transition distribution for the $\latent$s,
denoted $\prob(\latent_{\nodeIdx} \given \latent_{\nodeIdx -1})$. The joint distribution over the elements produced by Algorithm \ref{genproc} is
\begin{align}
\label{depstructure}
 \prob(\node_1, \latent_0) \prod_{\nodeIdx=1}^{\nodeLen} 
    \prob(\children_{\nodeIdx} \given \node_{\nodeIdx}, \latent_{\nodeIdx}) 
    \prob(\latent_{\nodeIdx} \given \latent_{\nodeIdx -1})
\end{align}
Thus, \OurModelShort{s} can be viewed as a Markov
model equipped with a stack---a special case of a Probabilistic Pushdown Automata (PPDA)
\cite{abney1999relating}.  Because depth-first order produces tokens in the same order that they are observed in the code, it is particularly well-suited. We note that other traversal orders produce valid distributions over trees such as right-left or
breadth-first. Because we compare to sequential models, we consider only depth-first.

 \begin{figure}[t]
 	\begin{center}
        \subfigure[pop $\node_1$, sample $\latent_1$]{
                \includegraphics[width=0.18\textwidth]{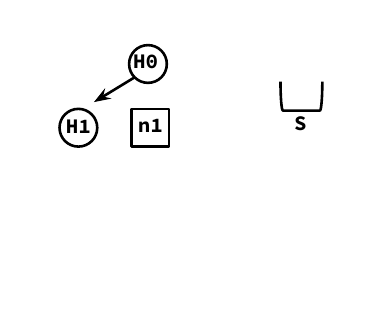}}
	~
        \subfigure[sample children, and push left-right]{
                \includegraphics[width=0.18\textwidth]{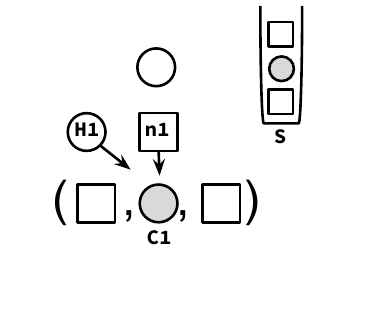}}
	~
        \subfigure[pop $\node_2$, sample $\latent_2$]{
                \includegraphics[width=0.2\textwidth]{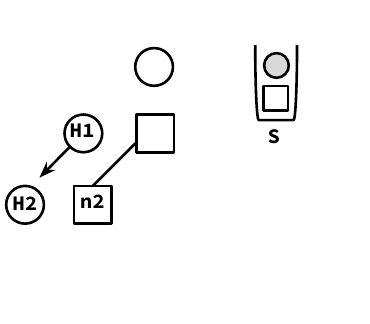}}
	~
        \subfigure[sample children tuple, and push left-right]{
                \includegraphics[width=0.2\textwidth]{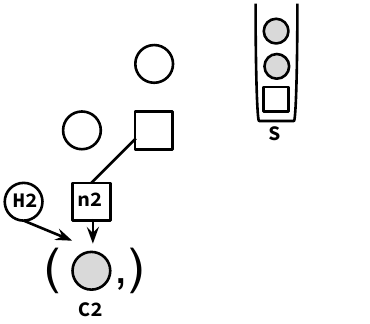}}
	~
        \subfigure[pop $\token_1, \token_2,$ and $\node_3$, sample $\latent_3$]{
                \includegraphics[width=0.2\textwidth]{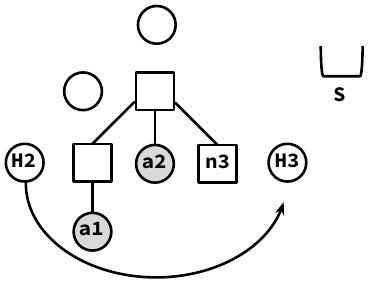}}
	\qquad
        \subfigure[sample children tuple, and push left-right]{
                \includegraphics[width=0.2\textwidth]{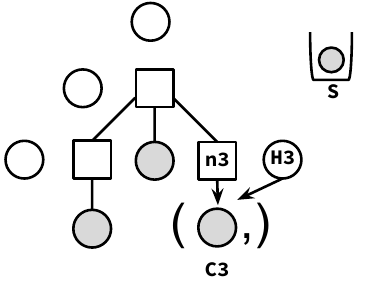}}
%	\qquad
%	        \subfigure[pop $\token_3$ for final state]{
%                \includegraphics[width=0.2\textwidth]{sample_stack_6.pdf}}
         	\end{center}	
        \caption{
Example sampling run from an \OurModelShort{}. Rectangles are
internal nodes, shaded circles are tokens, circles are
traversal variables, and stack $\stack$ is shown in the state
after the computations described in subcaption. Parentheses
indicate tuples of nodes and arrows indicate conditioning. Popping of
tokens omitted for brevity, but note that they are labelled in the
order encountered. 
}
\label{fig:samples}
\end{figure}

%%%%%%%%%%%%%%%%%%%%%%%%%%%%%%%%%%%%%%%%%%%%%%%%%%%%%%%%%%%%%%%%%%%%%%%%

%\vspace{-5pt}
%\subsection{Parameterizations}
%\label{subsec:logbilinear_parameterization}
%\vspace{-5pt}
{\bf Parameterizations. } Most of the uncertainty in generation comes
from generating children tuples. In order to avoid an explosion in the
number of parameters for the children tuple distribution
$\prob(\children \given \node, \latent)$, we use a log-bilinear form.
For all distributions other than $\prob(\children \given \node,
\latent)$ we use a simple tabular parameterization.

The log-bilinear form consists of a real-valued vector representation
of $(\node_{\nodeIdx}, \latent_{\nodeIdx})$ pairs,
$\repfn_{con}(\node_{\nodeIdx}, \latent_{\nodeIdx})$, a real-valued
vector representation for the children tuple,
$\repfn_{ch}(\children_{\nodeIdx})$, and a bias term for the children,
$\biasfn_{ch}(\children_{\nodeIdx})$.  These are combined via an inner
product, which gives the negative energy of the children tuple
\begin{align}
-\energy(\children_{\nodeIdx} ;  \node_{\nodeIdx},  \latent_{\nodeIdx}) & =
\repfn_{ch}(\children_{\nodeIdx}) ^{\!\trans} \repfn_{con}(\node_{\nodeIdx},  \latent_{\nodeIdx}) + \biasfn_{ch}(\children_{\nodeIdx}) \nonumber
\end{align}
As is standard, this is then
exponentiated and normalized to give the probability of sampling the children:
%\begin{align}
$\prob(\children_{\nodeIdx} \given \node_{\nodeIdx},  \latent_{\nodeIdx}) \propto
\exp \left\{ -\energy(\children_{\nodeIdx} ;  \node_{\nodeIdx},  \latent_{\nodeIdx}) \right\}.$
%\end{align}
%
We take the support over which to normalize this distribution to be
the set of children tuples observed as children of nodes of
type $\node_\nodeIdx$  in the training set.

The representation functions rely on the notion of an
$\representation$ matrix that can be indexed into with
objects to look up $\dims$ dimensional real-valued vectors. 
 $\representation_{x}$ denotes the row of the
$\representation$ matrix corresponding to any variable equal to $x$. For example, if $\node_{\nodeIdx} = \code{int}$ and $\node_j = \code{int}$, then $\representation_{\node_{\nodeIdx}} = \representation_{\node_{j}}$. These objects may be tuples and in particular $\code{(type, int)} \neq \code{int}$.
Similarly, $\bias_{x}$ looks up a real
number. 
In the simple variant, each unique $\children$ sequence
receives the representation
$\repfn_{ch}(\children_{\nodeIdx}) =
\representation_{\children_{\nodeIdx}}$ and
$\biasfn_{ch}(\children_{\nodeIdx}) = \bias_{\children_{\nodeIdx}}$.
The representations for $(\node, \latent)$ pairs are defined
as sums of representations of their components.
If $\latent_{\nodeIdx}$ is a collection of variables ($\latent_{\nodeIdx \latentIdx}$ representing the $\latentIdx$th
variable at the $\nodeIdx$th step) then
\begin{align}
\repfn_{con}(\node_{\nodeIdx},  \latent_{\nodeIdx}) & = \contextmatrixcon_0 \representation_{\node_{\nodeIdx}} + \sum_{\latentIdx=1}^\latentLen \contextmatrixcon_{\latentIdx} \representation_{\latent_{\nodeIdx \latentIdx}}
\end{align}
The $\contextmatrixcon$s are matrices (diagonal for computational
efficiency) that modulate the contribution of a variable in a
position-dependent way. In other variants the children tuple
representations will also be defined as sums of their component
representations.
The log-bilinear parameterization has the
desirable property that the number of parameters grows
linearly in the dimension of $\latent$, so we can afford to have high
dimensional traversal variables without worrying about exponentially
bad data fragmentation.

%%%%%%%%%%%%%%%%%%%%%%%%%%%%%%%%%%%%%%%%%%%%%%%%%%%%%%%%%%%%%%%%%%%%%%%%

%%%%%%%%%%%%%%%%%%%%%%%%%%%%%%%%%%%%%%%%%%%%%%%%%%%%%%%%%%%%%%%%%%%%%%%%%%%%%%%%%%%%%%%%%
%%%%%%%%%%%%%%%%%%%%%%%%%%%%%%%%%%%%%%%%%%%%%%%%%%%%%%%%%%%%%%%%%%%%%%%%%%%%%%%%%%%%%%%%%
\vspace{-5pt}
\section{Extending \OurModelShort s}
\vspace{-5pt}
\label{sec:extending}
The extensions of \OurModelShort{s} in this section allow 
(a) certain traversal variables to depend arbitrarily on  
previously generated elements of the AST;
%history 
(b) annotating nodes with richer types; and
(c) letting $\repfn_{ch}$ be compositionally defined, which becomes powerful
when combined with deterministic reasoning about variable scoping.

We distinguish
between \emph{\certain} and \emph{latent} traversal variables.  The
former can be computed deterministically from the current partial tree
(the tree nodes and tokens that have been instantiated at step
$\nodeIdx$) that has been generated while the latter cannot. To refer to
a collection of both \certain~and latent traversal variables we continue to use the unqualified ``traversal variables'' term.

%\commentout{
%By generating the AST we liberate ourselves to
%condition on knowledge of the incomplete tree when sampling
%children. In particular, we are free to employ compiler-like
%reasoning on the incomplete tree. Because we can treat ASTs as
%observed, these extension do not cost the model in
%efficiency. Moreover, the independence assumptions encoded in
%Algorithm \ref{genproc} allow us target our efforts. By identifying
%parent nodes whose children distribution have particularly high
%entropy and constructing ``smarter'' children distributions we can
%improve the model in a modular fashion.
%}
%
%\vspace{-5pt}
%\subsection{Deterministic Traversal Variables}
%\vspace{-5pt}

{\bf Deterministic Traversal Variables. } In the basic generative procedure,
traversal
variables $\latent_{\nodeIdx}$ satisfy the first-order Markov
property, but it is possible to condition on any
part of the tree that has already been produced. That is, we can replace $\prob(\latent_{\nodeIdx} \given \latent_{\nodeIdx -1})$ by
$\prob(\latent_{\nodeIdx} \given \latent_{0:\nodeIdx -1}, \node_{1:\nodeIdx}, \token_{1:\tokenIdx})$ in \eqref{depstructure}.
Inference becomes complicated unless
these variables are \certain~traversal variables (inference is explained in \secref{sec:inference_and_learning}) and the unique
value that has support can be computed efficiently. %\begin{align}
%\prob(\latent_{\nodeIdx} \given \latent_{\nodeIdx -1}) \longrightarrow 
%\end{align}
Examples of these variables include the set of node types 
that are ancestors of a given node, and the last $n$ tokens
or internal nodes that have been generated.
%We distinguish variables that are
%hierarchical (functions of the path to the root of the AST) and
%sequential (functions of some fixed window of nodes or tokens just
%produced). 
Variable scoping, a more elaborate deterministic
relationship, is considered later.

%\vspace{-5pt}
%\subsection{Annotating Nodes}
%\vspace{-5pt}
%\label{sec:annotating}
%
{\bf Annotating Nodes. } Other useful features may not be deterministically computable from the
current partial tree. Consider knowing that a
\code{BinaryExpression} will evaluate to an object of type
\code{int}. This information can be encoded by
letting nodes take values in the cross-product space of the node type
space and the annotation space.  For example, when adding type
annotations we might have nodes take value (\code{BinaryExpression}, \code{int}) where before they were just \code{BinaryExpression}.
This can be problematic, because the cardinality of the parent node space increases exponentially as we add annotations. Because
the annotations are uncertain, this means there are more choices of node values
at each step of the generative procedure, and this incurs a cost in
log probabilities when evaluating a model.  Experimentally we found that
simply annotating expression nodes with type information led to worse
log probabilities of generating held out data:\ the cost of generating
tokens decreased because the model had access to type information, the
increased cost of generating type annotations along with nodetypes outweighed
the improvement. 

%\TODO{Nicolas Heess suggested to me that this behavior is
%an artifact of our suboptimal inference, and that if we properly integrated
%over all trees that could have generated a program, this behavior wouldn't
%arise. Maybe we should downplay this a bit.}

%This can lead to
%a fragmentation of the dataset, but this can ameliorated with
%log-bilinear parameterizations. Worse, if the annotations are not
%deterministically propagated, then the models pays some cost in
%predicting them.

%\vspace{-5pt}
%\subsection{Identifier Token Scoping}
%\vspace{-5pt}
%\label{sec:scoping}

{\bf Identifier Token Scoping. } The source of greatest uncertainty when generating a program are children of \code{IdentifierToken}
nodes.
\code{IdentifierToken} nodes are very common and are parents of
all tokens (e.g. variable and method
names) that are not built-in language keywords (e.g.,
\code{IntKeyword} or \code{EqualsToken}) or constants (e.g.,
\code{StringLiteral}s).  Knowing which variables have previously been declared and are currently in
scope is one of the most powerful signals when predicting which
\code{IdentifierToken} will be used at any point in a program. Other useful cues include how recently the
variable was declared and what the type the variable is. In this section we a scope model for \OurModelShort{s}.

Scope can be represented as a set of variable feature
vectors corresponding to each to a variable that is in scope.\footnote{Technically, we view the scope as a deterministic traversal variable, but it does not contribute to $\repfn_{con}$.}  Thus, each feature vector contains a string identifier corresponding
to the variable along with other features as (key, value) tuples, for example $(\code{type}, \code{int})$.
A variable is ``in scope'' if there is a feature vector in the scope
set that has a string identifier that is the same as the variable's
identifier.

%\commentout{
%One of the most powerful properties of the AST is that it completely
%defines the scope of every identifier token. Knowing which identifier
%tokens are in scope limits the support by restricting the possible
%identifiers to those corresponding to previously declared
%variables. Moreover, when a variable is declared and put into scope
%there is information available that affects its use later in a
%program. A scope is a convenient way to carry this information and the
%information about which identifier tokens are valid. In order to
%condition on the scope when choosing identifier tokens we need to be
%careful of two things. First, that we are required to predict when to
%sample from the local scope and when to sample a global
%variable. Second, we have to ensure that the scope is well defined at
%every $\node_{\nodeIdx}$ and that we do not put things in scope or
%remove them without generating all the information necessary to make
%that decision. In particular we need to know when to add and when to
%remove variables. Luckily this information can all be computed from
%the AST.
%}

When sampling an identifier token, there is a two step procedure.
First, decide whether this identifier token will be sampled from the
current scope. This is accomplished by annotating each
\code{IdentifierToken} internal node with a binary variable that has
the states \code{global} or \code{local}.  If \code{local},
proceed to use the local scope model defined next.  If
\code{global}, sample from a global identifier token model that
gives support to all identifier tokens. Note, we consider the
\code{global} option a necessary smoothing device, although ideally
we would have a scope model complete enough to have all
possible identifier tokens.

The scope set can be updated deterministically as we traverse the
AST by recognizing patterns that correspond to when variables should
be added or removed from the scope.  We implemented this logic for three 
cases: parameters of a method, locally declared variables, 
and class fields that have been defined prior in the
class definition.  We do not include class fields defined after the
current point in the code, and variables and methods available in
included namespaces.  This incompleteness necessitates
the \code{global} option described above, but these three cases
are very common and cover many interesting cases.

%\commentout{
%, because we are restricted to the incomplete tree
%generated in a left-to-right depth-first manner. Because of this we
%can think of scope as a discrete variable $\latent_{\nodeIdx
%\latentIdx}$ that represents an element of the power set of identifier
%tokens. Adding and removing identifier tokens is simply a matter of
%propagating information down the AST that tracks where an identifier
%token \emph{would} pass out of scope \emph{if} it were declared at
%this point. We can annotate the pusher, which is the parents next
%rightmost sibling, and the popper varies depending on the declaration
%type, but is guaranteed to be on the stack because it is the forward
%sibling of some ancestor.
%\TODO{IS THIS EXPLANATION ENOUGH? STILL FEEL SKETCHED OUT BY SCOPE}
%\TODO{What is done when you're in the test set and you're asked to push a variable onto the scope but you've never seen that variable before? Do you they get Rep vectors that are 0? Does this influence the support or the logp?}
%}
%

Given the scope set which contains variable feature vectors 
$\{\varfeat_{\token}\}$ and parent node \code{(IdentifierToken, local)}, 
the probability of selecting token child $\token$ is proportional to
%\begin{align}
$\prob(\token \given \node_{\nodeIdx},  \latent_{\nodeIdx}) \propto
\exp \left\{ -\energy(\token ;  \node_{\nodeIdx},  \latent_{\nodeIdx}) \right\}$,
%\end{align}
where we normalize only over the variables currently in scope.
%, and where
%the representation of the possible children is defined compositionally.
Specifically, we let
$\repfn_{ch}(\token)$ and $\biasfn_{ch}(\token)$ be defined as follows:
\begin{align}
\repfn_{ch}(\token) & = \sum_{\varfeatIdx=1}^\varfeatLen \contextmatrixch_{\varfeatIdx} \representation_{\varfeat_{\token\varfeatIdx}} \;\;\;\; \biasfn_{ch}(\token) = \sum_{\varfeatIdx=1}^\varfeatLen \bias_{\varfeat_{\token\varfeatIdx}}.
\end{align}
For example, if a variable in scope has feature vector 
(\code{numNodes}, (\code{type}, \code{int}) , (\code{recently-declared}, \code{0})), then its corresponding $\repfn_{ch}$ would be
a context matrix-modulated sum of representations $\representation_{\code{numNodes}}$,
$\representation_{(\code{type}, \code{int})}$, and $\representation_{(\code{recently-declared}, \code{0})}$.
This representation will then be combined with the context representation
as in the basic model. The string identifier feature $\code{numNodes}$ is the same object as token nodes of the same string, 
thus they share their representation.

%%%%%%%%%%%%%%%%%%%%%%%%%%%%%%%%%%%%%%%%%%%%%%%%%%%%%%%%%%%%%%%%%%%%%%%%%%%%%%%%%%%%%%%%%
%%%%%%%%%%%%%%%%%%%%%%%%%%%%%%%%%%%%%%%%%%%%%%%%%%%%%%%%%%%%%%%%%%%%%%%%%%%%%%%%%%%%%%%%%
%\commentout{
%\subsection{Other Specialized Models}
%\TODO{Talk about how we can go in and modify a model by nodetype 
% or by other subset of parents.  Give example of BlockSyntax.
%Then say we observed that a big logprob cost came from IdentifierTokens,
%so we pushed further on this model, incorporating scope information.}
%
%\TODO{probably cut, but we could mention other possibilities}
%\begin{itemize}
%\item Additional annotations that we use.
%\item Give example of BlockSyntax model, where we first sample number of statements,
%then sample the statements.
%\item Maybe one other example.  Depth-specific BinaryExpression?
%\end{itemize}
%}

%%%%%%%%%%%%%%%%%%%%%%%%%%%%%%%%%%%%%%%%%%%%%%%%%%%%%%%%%%%%%%%%%%%%%%%%%%%%%%%%%%%%%%%%%
%%%%%%%%%%%%%%%%%%%%%%%%%%%%%%%%%%%%%%%%%%%%%%%%%%%%%%%%%%%%%%%%%%%%%%%%%%%%%%%%%%%%%%%%%

\vspace{-5pt}
\section{Inference and Learning in \OurModelShort s}
\vspace{-5pt}
\label{sec:inference_and_learning}

In this section we briefly consider how to compute gradients and probabilities in \OurModelShort{s}.

\commentout{
At this point
it is useful to distinguish between two cases.  In the first case,
all traversal variables are \certain, and computation will
be very straightforward.  In the second case, latent traversal
variables are allowed, and we will need to use dynamic programming
to compute log probabilities and expectation maximization (EM) for
learning.
}

{\bf Only \Certain~Traversal Variables. }
If all traversal variables $\latent_{\nodeIdx}$ can be computed deterministically from the 
current partial tree, we use the compiler to compute the full AST corresponding
to program $\tokens_m$. From the AST we compute the only valid
setting of the traversal variables.  Because both the AST and the traversal variables can be deterministically computed from the token sequence, all variables in the model 
can be treated as observed.  Since \OurModelShort s are directed models, this means that 
the total log probability is a sum of log probabilities at each production, and 
learning decomposes into independent problems at each production.  Thus, we can simply stack
all productions into a single training set and follow standard gradient-based procedures for
training log-bilinear models.  More details will be described in \secref{sec:experiments}, 
but generally
we follow the Noise-Contrastive Estimation (NCE) technique employed in \citet{MnihTeh2012}.

{\bf Latent Traversal Variables. } In the second case, we allow latent
traversal variables that are not deterministically computable from the AST.
In this case, the traversal
variables couple the learning across different productions from the
same tree.  For simplicity and to allow efficient exact inference, we
restrict these latent traversal variables to just be a single discrete
variable at each step (although this restriction could easily be
lifted if one was willing to use approximate inference).  Because the
AST is still a deterministic function of the tokens, computing log
probabilities corresponds to running the forward-backward algorithm over the
latent states in the depth-first traversal of the AST. We can
formulate an EM algorithm adapted to the NCE-based learning of
log-bilinear models for learning parameters. The details of this can
be found in the Supplementary Material.

\vspace{-5pt}
\section{Related Work}
\vspace{-5pt}
\label{related}
The \OurModelShort{}s described here are closely related to several
existing models.
Firstly, a Hidden Markov Model (HMM) can be recovered by having all children
tuples contain a token and a \code{Next} node, or just a token (which will terminate
the sequence), and having a single discrete latent traversal variable.
If the traversal variable has only one state and the children
distributions all have finite support, then an \OurModelShort{}
becomes equivalent to a Probabilistic Context Free Grammar (PCFG).
% in
%which the non-terminal tokens are the internal nodes and the terminal
%tokens are the tokens. 
PCFGs and their variants are components of
state-of-the-art parsers of English
\cite{mcclosky2006effective}, and many variants have been explored: internal node annotation \citet{charniak1997statistical} and latent annotations \citet{matsuzaki2005probabilistic}.
Aside from the question of the order of the traversals, the traversal variables make \OurModelShort{s} special cases of Probabilistic
Pushdown Automata (PPDA) (for definition and weak equivalence to PCFGs, see
\citet{abney1999relating}).  
Log-bilinear parameterizations have been applied widely in language
modeling, for \ngram~models \cite{saul1997aggregate, MnihHinton2007, MnihTeh2012}
and PCFG models \cite{charniak2000maxentparse, klein2002fast, titov2007incrbayesnets, henderson2010incrsigbeliefnets}.
To be clear, our claim is not that general
Tree Traversal models or the log-bilinear paremeterizations are
novel; however, we believe that the full \OurModelShort~construction,
including the tree traversal structure, log-bilinear parameterization, and
incorporation of deterministic logic to be novel and of general interest.

\commentout{
It is true in general that PPDAs and PCFGs
are equivalent classes of distributions over \emph{terminal} tokens,
although they are subject to different inductive biases
\cite{abney1999relating}. Yet, if the traversal variables of
\OurModelShort{}s are latent and marginalized, then the resulting
model is \emph{not} context free with respect to the tree.
}

\commentout{
This
approach can be seen as a sort of low-rank approximation to
parameterizing distributions with a single number, an approach that
enjoys a large literature \cite{SalMnih08, MnihHinton2007}. Approaches that reduce the number of
parameters with low-rank approximations in $n$-gram models deal
effectively with analogous fragmentation issues
. In fact, log-linear
parameterizations have been used in PCFGs
\cite{charniak2000maxentparse} and factorized models that ameliorate
fragmentation in similar ways to low-rank parameterizations have also been
explored \cite{klein2002fast}.
}

The problem of modeling source code is relatively
understudied in machine learning.  We previously mentioned 
\citet{hindle2012naturalness} and \citet{allamanis2013mining}, which
tackle the same task as us but with simple NLP models. 
Very recently, \citet{allamanis2014mining} explores more sophisticated
nonparametric Bayesian grammar models of source code for the purpose of
learning code idioms.
%There is also quite a bit of work on mining frequent API usage patterns for
%use in suggesting method calls on a variable and program point
%currently under consideration.  These approaches use some amount of
%basic machine learning, typically following the approach of defining a
%similarity measure between code contexts, then re-ranking candidate
%method calls based on frequency \cite{bruch2009learning,nguyen2012graphbased,
%wang2013mining}.
\citet{liang10programs}
use a sophisticated non-parametric model to encode the prior that
programs should factorize repeated computation, but there is no
learning from existing source code, and the prior is only
applicable to a functional programming language with quite simple syntax rules. 
Our approach builds a sophisticated and learned model
and supports the full language specification of
a widely used imperative programming language.

%%%%%%%%%%%%%%%%%%%%%%%%%%%%%%%%%%%%%%%%%%%%%%%%%%%%%%%%%%%%%%%%%%%%%%%%%%%%%%%%%%%%%%%%%
%%%%%%%%%%%%%%%%%%%%%%%%%%%%%%%%%%%%%%%%%%%%%%%%%%%%%%%%%%%%%%%%%%%%%%%%%%%%%%%%%%%%%%%%%

\vspace{-5pt}
\section{Experimental Analysis}
\vspace{-5pt}
\label{sec:experiments}

In all experiments, we used a dataset that we collected from
TopCoder.com.  There are 2261 C\# programs which make up 140k lines of
code and 2.4M parent nodes in the collective abstract syntax trees.
These programs are solutions to programming competitions, and there is
some overlap in programmers and in problems across the programs.  We
created training splits based on the user identity, so the set of
users in the test set are disjoint from those in the training or
validation sets (but the training and validation sets share users).
The overall split proportions are 20\% test, 10\%
validation, and 70\% train.
The evaluation measure that we use throughout is the log probability
under the model of generating the full program.  All logs are base 2.
To make this number more easily interpretable, we divide by the number of tokens in 
each program, and report the average log probability per token.

\textbf{Experimental Protocol. }
All experiments use a validation set to choose
hyperparameter values.  These include the strength of a smoothing
parameter and the epoch at which to stop training (if applicable).  We
did a coarse grid search in each of these parameters and the numbers
we report (for train, validation, and test) are all for the settings
that optimized validation performance.  For the gradient-based
optimization, we used AdaGrad \cite{duchi2011adaptive} with stochastic
minibatches.  Unless otherwise specified, the dimension of the latent
representation vectors was set to 50.  Occasionally the test set will
have tokens or children tuples unobserved in the training set.  In
order to avoid assigning zero probability to the test set, we locally
smoothed every children distribution with a default model that gives
support to all children tuples. The numbers we report are a lower bound on the log probability of data under for a
mixture of our models with this default model.  Details of this
smoothed model, along with additional experimental details, appear in
the Supplementary Materials. There is an additional question of how to
represent novel identifiers in the scope model. We set the
representations of all features in the variable feature vectors that
were unobserved in the training set to the all zeros vector.

{\bf Baselines and Basic Log-bilinear Models. }
The natural choices for baseline models are \ngram~models and
PCFG-like models.  In the \ngram~models we use additive smoothing,
with the strength of the smoothing hyperparameter chosen to optimize validation
set performance.  Similarly, there is a smoothing
parameter in the PCFG-like models that is chosen to optimize validation
set performance.  
We explored the effect of the log-bilinear parameterization in
two ways.  First, we trained a PCFG model that was identical to the
first PCFG model but with the parameterization defined using the
standard log-bilinear parameterization.  This is the most basic \OurModelShort~model,
with no traversal variables (\OurModelShort-$\emptyset$).
The result was nearly identical to the standard PCFG.  Next, we trained a 10-gram
model with a standard log-bilinear parameterization, which is equivalent to the models
discussed in \cite{MnihTeh2012}.  This approach dominates the basic
\ngram~models, allowing longer contexts and generalizing better.
Results appear in \figref{fig:baselines}.

\begin{figure}[t]
\centering
\footnotesize{
\begin{tabular}{|r|ccc|}
\hline
Method & Train & Valid & Test  \\
\hline\hline
2-gram &	-4.28 &	-4.98 &	-5.13 \\
3-gram & 	-2.94 &	-5.05 &	-5.25 \\
4-gram & 	-2.70 &	-6.05 &	-6.31 \\
5-gram & 	-2.68 &	-7.22 &	-7.45 \\
\hline
PCFG   & 	-3.67 &	-4.04 & -4.23 \\
\hline
\OurModelShort-$\emptyset$ & -3.67	&	-4.04 & -4.24  \\
LBL 10-gram & -3.19 & -4.62 & -4.87 \\
\hline
\end{tabular}
}
\caption{Baseline model log probabilities per token.}
\label{fig:baselines}
\end{figure}
\begin{figure}[t]
\centering
\footnotesize{
\begin{tabular}{|r|ccc|}
\hline
Method & Train & Valid & Test  \\
\hline\hline
\OurModelShort-$\emptyset$ & -3.67	&	-4.04 & -4.24  \\
\hline
\OurModelShort-Seq     &-2.54  &-3.25 &-3.46  \\
\OurModelShort-Hi    &-2.28  &-3.30	&-3.53	 \\
\OurModelShort-HiSeq &-2.10  &-3.06	&-3.28 	 \\
\hline
\end{tabular}
}
\caption{\OurModelShort~models augmented with determistically
determinable latent variables.}
\label{fig:hiandseq}
\end{figure}

{\bf Deterministic Traversal Variables. } Next, we augmented \OurModelShort-$\emptyset$ model with \certain~traversal
variables that include hierarchical and sequential information.  
The hierarchical information is
the depth of a node, the kind of a node's parent, and a
sequence of 10 ancestor history variables, which store for the last 10
ancestors, the kind of the node and the index of the child that would need
to be recursed upon to reach the current point in the tree.
The sequential information is the last 10 tokens that were generated.

In \figref{fig:hiandseq} we report results for three variants:\ 
hierarchy only (\OurModelShort-Hi),
sequence only (\OurModelShort-Seq), and both (\OurModelShort-HiSeq). 
The hierarchy features alone perform better than the sequence features alone, but
that their contributions are independent enough that the combination
of the two provides a substantial gain over either of the individuals.

\begin{figure}[t]
\centering
\footnotesize{
\begin{tabular}{|r|ccc|}
\hline
Method & Train & Valid & Test  \\
\hline\hline
%Unigram & TODO & TODO & TODO \\
\OurModelShort-$\emptyset$ & -3.67	&	-4.04 & -4.24  \\
\hline
\OurModelShort-latent & -3.23 &	-3.70 & -3.91  \\
LBL HMM & -9.61 & -9.97 & -10.10 \\
\hline
\end{tabular}
}
\caption{\OurModelShort-latent models augmented with latent
variables and learned with EM.}
\label{fig:latent}
\end{figure}

\begin{figure}[t]
\centering
\footnotesize{
\begin{tabular}{|r|ccc|}
\hline
Method & Train & Valid & Test  \\
\hline\hline
\OurModelShort-HiSeq (50) &-2.10  &-3.06	&-3.28 	 \\
\hline
\OurModelShort-HiSeq-Scope (2) &-2.28  &-2.65   &-2.78 \\
\OurModelShort-HiSeq-Scope (10) &-1.83  &-2.29  &-2.44  \\
\OurModelShort-HiSeq-Scope (50) &-1.54  &-2.18  &-2.33 \\
\OurModelShort-HiSeq-Scope (200) &-1.48  &-2.16  &-2.31  \\
\OurModelShort-HiSeq-Scope (500) &-1.51  &-2.16  & -2.32\\
\hline
\end{tabular}
}
\caption{\OurModelShort~models with \certain~traversal variables and scope
 model.  Number in parenthesis is
the dimension of the representation vectors.}
\label{fig:scopes}
\end{figure}

{\bf Latent Traversal Variables. } Next, we considered latent traversal variable \OurModelShort{} models trained with EM learning.
In all cases, we used 32 discrete latent states.  Here, results were more mixed.
While the latent-augmented \OurModelShort~(\OurModelShort-latent) outperforms the \OurModelShort-$\emptyset$
model, the gains are smaller than achieved by adding the deterministic features.
As a baseline, we also trained a log-bilinear-parameterized standard HMM, and
found its performance to be far worse than other models.
We also tried a variant where we added latent traversal variables to 
the \OurModelShort-HiSeq model from the previous section, but 
the training was too slow to be practical due to the cost of computing
normalizing constants in the E step.  See \figref{fig:latent}.
%We gave
%the learning a runtime budget of a week, but after this time, it had not
%surpassed the performance of the non-latent version.

{\bf Scope Model. } The final set of models that we trained incorporate the scope
model from \secref{sec:extending} (\OurModelShort-HiSeq-Scope). The features of variables that we
use are the identifier string, the type, where the variable
appears in a list sorted by when the variable was declared
(also known as a de Bruijn index), and where the variable
appears in a list sorted by when the variable was last assigned
a value.  Here, the additional structure provides
a large additional improvement over the previous best model
(\OurModelShort-HiSeq).  See \figref{fig:scopes}.

{\bf Analysis. } To understand better where the improvements in the different models come from,
and to understand where there is still room left for improvement in the
models, we break down the log probabilities from the previous experiments
based on the value of the parent node.  The results appear in
\figref{fig:tree_token_breakdowns}.  In the first column is the total log probability
number reported previously.  In the next columns, the contribution is split
into the cost incurred by generating tokens and trees respectively.
We see, for example, that the full scope model pays a slightly higher cost
to generate the tree structure than the Hi\&Seq model, which is due to it having
to properly choose whether IdentifierTokens are drawn from local or global
scopes, but that it makes up for this by paying a much smaller cost when
it comes to generating the actual tokens.

In the Supplementary Materials, we go further into the breakdowns
for the best performing model, reporting the percentage of total cost
that comes from the top parent kinds.  \code{IdentifierToken}s from the
global scope are the largest cost (30.1\%), with  \code{IdentifierToken}s
covered by our local scope model (10.9\%) and \code{Block}s (10.6\%) next.
This suggests that there would be value in extending our scope
model to include more \code{IdentifierToken}s and an improved model of Block sequences.

\begin{figure}
\centering
\footnotesize{
\begin{tabular}{|r|ccc|}
\hline
Method & Total & Token & Tree  \\
\hline\hline
\OurModelShort-$\emptyset$ & -4.23 &	-2.78 &	-1.45 \\
\OurModelShort-Seq &-3.53&	-2.28&	-1.25\\
\OurModelShort-Hi &-3.46&	-2.18&	-1.28\\
\OurModelShort-HiSeq &-3.28&	-2.08&	-1.20\\
\OurModelShort-HiSeq-Scope & -2.33&	-1.10&	-1.23\\
\hline
\end{tabular}
}
\caption{
Breakdowns of test log probabilities by whether the
cost came from generating the tree structure or tokens.
For all models $D=50$.
}
\label{fig:tree_token_breakdowns}
\vspace{-10pt}
\end{figure}

{\bf Samples. } Finally, we qualitatively evaluate the different methods by drawing samples
from the models.  Samples of \code{for} loops
appear in \figref{fig:intro_fig}.  
To generate these samples, we ask (b) the PCFG and (c) the \OurModelShort-HiSeq-Scope
model to generate a \code{ForStatement}.  For (a) the LBL \ngram~model, we simply insert
a \code{for} token as the most recent token.  We also initialize the traversal variables
to reasonable values:\ e.g., for the \OurModelShort-HiSeq-Scope~model, we initialize the local scope to
include \code{string[] words}.  We also provide samples 
of full source code files (\code{CompilationUnit}) from the \OurModelShort-HiSeq-Scope~model in the Supplementary Material, and additional \code{for} loops. Notice the 
structure that the model is able to capture, particularly related to 
high level organization, and variable use and re-use.  It also learns quite subtle things, like \code{int} variables
often appear inside square brackets.

%%%%%%%%%%%%%%%%%%%%%%%%%%%%%%%%%%%%%%%%%%%%%%%%%%%%%%%%%%%%%%%%%%%%%%%%%%%%%%%%%%%%%%%%%
%%%%%%%%%%%%%%%%%%%%%%%%%%%%%%%%%%%%%%%%%%%%%%%%%%%%%%%%%%%%%%%%%%%%%%%%%%%%%%%%%%%%%%%%%
%%%%%%%%%%%%%%%%%%%%%%%%%%%%%%%%%%%%%%%%%%%%%%%%%%%%%%%%%%%%%%%%%%%%%%%%%%%%%%%%%%%%%%%%%
%%%%%%%%%%%%%%%%%%%%%%%%%%%%%%%%%%%%%%%%%%%%%%%%%%%%%%%%%%%%%%%%%%%%%%%%%%%%%%%%%%%%%%%%%

\vspace{-5pt}
\section{Discussion}
\vspace{-5pt}

Natural source code is a highly structured source of data that has
been largely unexplored by the machine learning community. We have built
probabilistic models that capture some of the structure that appears
in NSC.  A key to our approach is to leverage the great deal of work
that has gone into building compilers.  The result is models that not
only yield large improvements in quantitative measures over baselines,
but also qualitatively produce far more realistic samples.

There are many remaining modeling challenges.  One question is how to
encode the notion that the point of source code is to \emph{do
something}.  Relatedly, how do we represent and discover high level
structure related to trying to accomplish such tasks?  There are also a
number of specific sub-problems that are ripe for further study.  Our
model of \code{Block} statements is naive, and we see that it
is a significant contributor to log probabilities.  It would be
interesting to apply more sophisticated sequence models to children
tuples of \code{Block}s. Also, applying the compositional
representation used in our scope model to other children tuples would interesting.  Similarly, it would be interesting to
extend our scope model to handle method calls.  Another high level
piece of structure that we only briefly experimented with is type
information.  We believe there to be great potential in properly
handling typing rules, but we found that the simple approach of
annotating nodes to actually hurt our models.

More generally, this work's focus was on generative modeling.
An observation that has become popular in machine learning lately is that learning
good generative models can be valuable when the goal is to
 extract features from the data. It would be interesting
to explore how this might be applied in the case of NSC.

In sum, we argue that probabilistic modeling of source code provides
a rich source of problems with the potential to drive forward new
machine learning research, and we hope that this work helps
illustrate how that research might proceed forward.

\vspace{-5pt}
\section*{Acknowledgments} 
\vspace{-5pt}
We are grateful to John Winn, Andy Gordon, Tom Minka, and Thore Graepel for helpful discussions and suggestions. We thank Miltos Allamanis and Charles Sutton for pointers to related work.

\small{
\bibliography{biblio}
\bibliographystyle{icml2014}
}

\clearpage
\newpage

\section*{Supplementary Materials for ``Structured Generative Models of Natural Source Code''}

\section*{EM Learning for Latent Traversal Variable \OurModelShort{s}}

Here we describe EM learning of \OurModelShort{s} with latent traversal variables. Consider probability of $\tokens$
 with deterministic traversal variables $\latent_{\nodeIdx}^{d}$ and latent traversal variables $\oneLatent_{\nodeIdx}^{l}$ (where $\latent_{\nodeIdx}$
represents the union of $\{ \oneLatent_{\nodeIdx}^{l} \}$ and
$\latent_{\nodeIdx}^{d}$):
\begin{align}
 \sum_{\latents} & \prob(\node_1, \latent_0) \prod_{\nodeIdx=1}^{\nodeLen} 
    \prob(\children_{\nodeIdx} \given \node_{\nodeIdx}, \latent_{\nodeIdx}) 
    \prob(\oneLatent^{l}_{\nodeIdx} \given \oneLatent^{l}_{\nodeIdx -1}) \nonumber \\
&   \qquad  \times \prob(\latent^{d}_{\nodeIdx} \given \latent_{0:\nodeIdx -1}, \node_{1:\nodeIdx}, , \token_{1:\tokenIdx})
\end{align}
Firstly, the $\prob(\latent_{\nodeIdx}^{d} \given \cdot)$ terms drop off because
as above we can use the compiler to compute the AST from $\tokens$ then use
the AST to deterministically fill in the only legal values for the $\latent_{\nodeIdx}^{d}$
variables, which makes these terms
always equal to 1.  It then becomes clear that 
the sum can be computed using the forward-backward algorithm.
For learning, we follow the standard EM formulation and lower bound the
data log probability
with a free energy of the following form (which for brevity drops the
prior and entropy terms):
\begin{align}
%\log \prob(\tokens) \ge 
%& \sum_{\latents} Q(\latents) \log \frac{\prob(\tokens, \AST_{int}, \latents)}{Q(\latents)} \\
  & \sum_{\nodeIdx=2}^N \sum_{\oneLatent^{l}_{\nodeIdx}, \oneLatent^{l}_{\nodeIdx-1}}  
    \!\! Q_{\nodeIdx,\nodeIdx-1}(\oneLatent^{l}_{\nodeIdx}, \oneLatent^{l}_{\nodeIdx-1}) \log P(\oneLatent^{l}_{\nodeIdx} \given \oneLatent^{l}_{\nodeIdx-1}) \nonumber \\
   & + \sum_{\nodeIdx=1}^N \sum_{\oneLatent^{l}_\nodeIdx} Q_\nodeIdx(\oneLatent^{l}_\nodeIdx) \log \prob(\children_\nodeIdx \given \node_\nodeIdx, \latent_\nodeIdx)
\end{align}
In the E step, the Q's are updated optimally given the current parameters
using the forward backward algorithm.
In the M step, given $Q$'s, the learning decomposes across productions. We
represent the transition probabilities using a simple tabular representation and use stochastic gradient updates. For the emission terms, it is again
straightforward to use standard log-bilinear model training. The only difference
from the previous case is that there are now $\hmmStateSize$ training examples
for each $\nodeIdx$, one for each possible value of $\oneLatent^{l}_i$, which
are weighted by their corresponding $Q_\nodeIdx(\oneLatent^{l}_i)$. A simple way
of handling this so that log-bilinear training methods can be used unmodified is
to sample $\oneLatent_\nodeIdx^{l}$ values from the corresponding
$Q_\nodeIdx(\cdot)$ distribution, then to add unweighted examples to the
training set with $\oneLatent_\nodeIdx^{l}$ values being given their sampled
value.  This can then be seen as a stochastic incremental M step.

\section*{More Experimental Protocol Details}

For all hyperparameters that were not validated over (such
as minibatch size, scale of the random initializations, and learning
rate), we chose a subsample of the training set and manually chose a
setting that did best at optimizing the training log probabilities.
For EM learning, we divided the data into databatches, which contained
10 full programs, ran forward-backward on the databatch, then created
a set of minibatches on which to do an incremental M step using
AdaGrad.  All parameters were then held
fixed throughout the experiments, with the exception that we
re-optimized the parameters for the learning that required EM, and we
scaled the learning rate when the latent dimension changed.  Our code
used properly vectorized Python for the gradient updates and a C++
implementation of the forward-backward algorithm but was otherwise not
particularly optimized.  Run times (on a single core) ranged from a
few hours to a couple days.

\section*{Smoothed Model}

In order to avoid assigning zero probability to the test set, we assumed
knowledge of the set of all possible tokens, as well as all possible
internal node types -- information available in the Roslyn
API. Nonetheless, because we specify distributions over tuples of
children there are tuples in the test set with no
support. Therefore we smooth every $\prob(\children_{\nodeIdx} \given
\latent_{\nodeIdx}, \node_{\nodeIdx})$ by mixing it with a default
distribution $\prob_{\default}(\children_{\nodeIdx} \given
\latent_{\nodeIdx}, \node_{\nodeIdx})$ over children that gives broad
support.
\begin{align}
\prob_{\pi}(\children_{\nodeIdx} &  \given \latent_{\nodeIdx}, \node_{\nodeIdx}) = \notag \\ 
&\pi \prob(\children_{\nodeIdx} \given \latent_{\nodeIdx}, \node_{\nodeIdx}) + (1-\pi) \prob_{\default}(\children_{\nodeIdx} \given \latent_{\nodeIdx}, \node_{\nodeIdx})
\end{align}
For distributions whose children are all 1-tuples of tokens, the
default model is an additively smoothed model of the empirical
distribution of tokens in the train set. For other distributions we
model the number of children in the tuple as a Poisson distribution,
then model the identity of the children independently (smoothed
additively).

This smoothing introduces trees other than the Roslyn AST with
positive support. This opens up the possibility that there are
multiple trees consistent with a given token sequence and we can no
longer compute $\log \prob(\tokens)$ in the manner discussed in
\secref{sec:inference_and_learning}.
Still we report the log-probability of the AST, which
is now a lower bound on $\log \prob(\tokens)$.

\begin{figure}[h]
\centering
\footnotesize{
\begin{tabular}{|rcc|}
\hline
Parent Kind & \% Log prob & Count \\
\hline\hline
(\code{IdentifierToken}, \code{global}) &  30.1 &  17518 \\
(\code{IdentifierToken}, \code{local})  &   10.9 &  27600 \\
\code{Block} &   10.6 &  3556 \\
\code{NumericLiteralToken} &      4.3 &   8070 \\
\code{Argument} &        3.6 &   10004 \\
\code{PredefinedType} &  3.0 &   7890 \\
\code{IfStatement} &     2.9 &   2204 \\
\code{AssignExpression} &        2.4 &   2747 \\
\code{ExpressionStatement} &     2.1 &   4141 \\
\code{EqualsValueClause} &       2.0 &   3937 \\
\code{StringLiteralToken} &       1.9 &   680 \\
\code{AddExpression} &   1.9 &   1882 \\
\code{ForStatement} &    1.6 &   1759 \\
\hline
\end{tabular}
}
\caption{
Percent of log probability contributions coming from 
top parent kinds for \OurModelShort-HiSeq-Scope (50) model on test set.
}
\end{figure}

\begin{figure*}
\begin{verbatim}
for ( int i = 0 ; i < words . Length ; ++ i ) i = i . Replace ( "X" , i ) ;

for ( int j = 0 ; j < words . X ; j ++ ) {
 if ( j [ j ] == - 1 ) continue ;
 if ( words [ j ] != words [ j ] ) j += thisMincost ( 1 ) ;
 else {
   j = ( j + 1 ) % 2 ;
   words [ j + 1 ] += words [ 0 ] ;
 }
}

for ( int j = words ; j < words . Pair ; ++ j )
 for ( int i = 0 ; i < words . Length ; ++ i ) {
   isUpper ( i , i ) ;
 }

for ( int i = 0 ; i < words . Length ; ++ i ) {
 words [ i , i ] = words . Replace ( "*" , i * 3 ) ;
}

for ( int j = 360 ; j < j ; ) {
 if ( ! words . ContainsKey ( j ) ) {
   if ( words . at + " " + j == j ) return ume ( j , j ) ;
 } else {
   j = 100 ;
 }
}

for ( int c = 0 ; c < c ; ++ c )
 for ( int i = 0 ; i < c ; i ++ ) {
   if ( ! words [ i ] ) i = i ;
 }

for ( int i = 0 ; i < words . Length ; i ++ ) {
 i . Parse ( i ) ;
}

for ( int i = words ; i < 360 ; ++ i ) {
 words [ i ] = words [ i ] ;
 i = 4 ;
}
\end{verbatim}
\caption{More example \code{for} loops generated by
\OurModelShort-HiSeq-Scope (50).
Whitespace edited to improve readability.}
\end{figure*}

\begin{figure*}
\begin{verbatim}
using System ;
using System . Collections . Text ;
using System . Text . Text ;
using System . Text . Specialized ;
using kp . Specialized ;
using System . Specialized . Specialized ;

public class MaxFlow
{
    public string maximalCost(int[] a, int b)
    {
        int xs = 0;
        int board = 0;
        int x = a;
        double tot = 100;
        for (int i = 0; i < xs; i++) {
            x = Math.mask(x);
        }
        for (int j = 0; j < x; j++) {
            int res = 0;
            if (res > 0) ++res;
            if (res == x) {
                return -1;
            }
        }
        for (int i = 0; i < a.Length; i++) {
            if (a[i - 2].Substring(board, b, xs, b, i)[i] == 'B') x = "NO";
            else if (i == 3) {
                if (i > 1) {
                    x = x.Abs();
                }
                else if (a[i] == 'Y') return "NO";
            }
            for (int j = 0; j < board; j++) if (j > 0) board[i] = j.Parse(3);
        }
        long[] x = new int[a.Count];
        int[] dp = board;
        for (int k = 0; k < 37; k++) {
            if (x.Contains < x.Length) {
                dp[b] = 1000000;
                tot += 1;
            }
            if (x[k, k] < k + 2) {
                dp = tot;
            }
        }
        return "GREEN";
    }
}
\end{verbatim}
\caption{Example CompilationUnit generated by \OurModelShort-HiSeq-Scope (50).
Whitespace edited to improve readability.}
\end{figure*}

\begin{figure*}
\begin{verbatim}
using System ;
using System . sampling . Specialized ;

public class AlternatingLane 
{
 public int count = 0 ;
 int dc = { 0 , 0 , 0 , 0 } ;
 double suma = count . Reverse ( count ) ;

 public int go ( int ID , int To , int grape , string [ ] next ) 
 {
  if ( suma == 1000 || next . StartsWith . To ( ID ) [ To ] == next [ To ] ) return ;
  if ( next [ next ] != - 1 ) {
   next [ To , next ] = 1010 ;
   Console . Add ( ) ;
  }
  for ( int i = 0 ; i < ( 1 << 10 ) ; i ++ ) {
   if ( i == dc ) NextPerm ( i ) ;
   else {
    count ++ ;
   }
  }
  return div ( next ) + 1 ;
 }

 string solve ( string [ ] board ) {
  if ( board [ count ] == '1' ) {
   return 10 ;
  }
  return suma ;
 }
}
\end{verbatim}
\caption{Example CompilationUnit generated by \OurModelShort-HiSeq-Scope (50).
Whitespace edited to improve readability.}
\end{figure*}

\begin{figure*}
\begin{verbatim}
using System ;
using System . Collections . Collections ;
using System . Collections . Text ;
using System . Text . Text ;

public class TheBlackJackDivTwo 
{
 int dp = 2510 ;
 int vx = new int [ ] { - 1 , 1 , 1 , 2 } ;
 int [ ] i ;
 int cs ;
 double xy ;
 int pack2 ;

 long getlen ( char tree ) {
  return new bool [ 2 ] ;
 }

 int getdist ( int a , int splitCost ) 
 {
  if ( ( ( 1 << ( a + vx ) ) + splitCost . Length + vx ) != 0 ) 
    i = splitCost * 20 + splitCost + ( splitCost * ( splitCost [ a ] - i [ splitCost ] ) ) ;
  int total = i [ a ] ;
  for ( int i = 0 ; i < pack2 ; i ++ ) {
   if ( can ( 0 , i , a , i , i ) ) {
    total [ a ] = true ;
    break ;
   }
   i [ i ] -= i [ a ] ;
   total = Math . CompareTo ( ) ;
   saiki1 ( a , vx , a , i ) ;
  }
  return total + 1 < a && splitCost < a ;
 }
}
\end{verbatim}
\caption{Example CompilationUnit generated by \OurModelShort-HiSeq-Scope (50).
Whitespace edited to improve readability.}
\end{figure*}

\end{document}